\documentclass[aps,pra,twocolumn,superscriptaddress,showpacs]{revtex4}
\usepackage{amsmath,float,graphicx,subfigure}
\begin{document}

\title{Effects of collisions against thermal impurities in the
dynamics of a trapped fermion gas } 

\author{P. Capuzzi} \author{P. Vignolo} \affiliation{INFM-NEST and
Classe di Scienze, Scuola Normale Superiore, I-56126 Pisa, Italy}
\author{F. Toschi} \author{S. Succi} \affiliation{Istituto per le
Applicazioni del Calcolo, CNR, Viale del Policlinico 137, I-00161
Roma, Italy} \author{M. P. Tosi} \affiliation{INFM-NEST and Classe di
Scienze, Scuola Normale Superiore, I-56126 Pisa, Italy}

\begin{abstract}
We present a theoretical study of the dynamical behavior of a gas made
of ultracold fermionic atoms, which during their motions can collide
with a much smaller number of thermal bosonic impurities. The atoms
are confined inside harmonic traps and the interactions between the
two species are treated as due to $s$-wave scattering with a negative
scattering length modeling the $^{40}$K-$^{87}$Rb fermion-boson
system. We set the fermions into motion by giving a small shift to
their trap center and examine two alternative types of initial
conditions, referring to (i) a close-to-equilibrium situation in which
the two species are at the same temperature (well below the Fermi
temperature and well above the Bose-Einstein condensation
temperature); and (ii) a far-from-equilibrium case in which the
impurities are given a Boltzmann distribution of momenta while the
fermions are at very low temperatures. The dynamics of the gas is
evaluated by the numerical solution of the Vlasov-Landau equations for
the one-body distribution functions, supported by some analytical
results on the collisional properties of a fermion gas. We find that
the trapped gaseous mixture is close to the collisionless regime for
values of the parameters corresponding to current experiments, but can
be driven towards a collisional regime even without increasing the
strength of the interactions, either by going over to heavier impurity
masses or by matching the width of the momentum distribution of the
impurities to the Fermi momentum of the fermion gas.
\end{abstract}

\maketitle  

\section{Introduction}

Trapped spin-polarized Fermi gases can be considered as
non-interacting at ultralow temperatures since $s$-wave collisions are
forbidden by the Pauli principle and higher-wave collisions are
negligible \cite{Minguzzi2004a}. The collisionality of such gases has
been increased by mixing it either with a fermion gas in a different
spin state \cite{DeMarco1999a, Gensemer2001a, Granade2002a,
Jochim2002a, Dieckmann2002a} or with a Bose-Einstein condensed gas of
bosonic atoms in numbers exceeding those of the fermions by a few
orders of magnitude
\cite{Truscott2001a,Schreck2001a,Roati2002a,Goldwin2002a,
Hadzibabic2002a}. However, the collision rate of a fermion-fermion
mixture at low temperatures is limited by the Pauli blocking of
collisions as a result of the occupation of final states around the
Fermi level and, in the case of a boson-fermion mixture, the
collisionality can be strongly diminished by superfluidity of the
condensate \cite{Timmermans1998a}. On the other hand, it has been
shown by Ferlaino {\em et al.\/} \cite{Ferlaino2003a} that the number
of collisions increases if the bosons are thermal and this can be
realized by diminishing the condensation temperature as for instance
can be achieved by lowering the number of bosons. It was previously
pointed out by Amoruso {\em et al.\/} \cite{Amoruso1999a} that the
presence of a small number of bosonic impurities could drastically
increase the collisionality of a Fermi gas, to the point of driving it
from the collisionless to the collisional regime. However, the Fermi
sphere is almost fully occupied in ultracold gases close to
equilibrium as are realized in actual experiments, and the Pauli
principle will in this case still limit the scattering against
impurities. The dependence of collisionality on the presence of
impurities thus needs detailed investigation.

In this work we study the collisional properties of an atomic Fermi
gas interacting with {\em a few\/} atomic impurities. The presence of
the impurities induces a damping of oscillatory motions of the gas and
a shift of its natural oscillation frequencies. We focus on the
collision and damping rates as functions of the concentration of
impurities and of temperature for system parameters corresponding to
$^{40}$K-$^{87}$Rb fermion-boson mixtures relevant to current
experiments \cite{Ferlaino2003a}. The effects of the momentum spread
of the distribution of the impurities on the collision rate is also
examined by both analytical and numerical means.

The paper is organized as follows. In Sec.\ \ref{sec:physmodel} we
introduce the physical system under study, while Sec.\
\ref{sec:collrate} discusses the collision rate and reports analytical
results in two limiting cases, the details being presented in an
Appendix. In Sec.\ \ref{sec:results} we carry out Vlasov-Landau
numerical simulations to examine the dynamics of the mixture. Finally,
Sec.\ \ref{sec:remarks} offers some concluding remarks.

\section{The physical model}
\label{sec:physmodel} 
The system that we study is a spin-polarized Fermi gas in a trap
containing bosonic impurities, which are free to move inside their own
trap and interact with the fermions both through a mean-field
potential and through collisions. The temperature of the system is
finite and well below the Fermi temperature $T_F$ , but well above the
critical temperature for Bose-Einstein condensation of the impurities.

The fermionic component ($j=F$) and the bosonic impurities ($j=B$) in
harmonic confining potentials $V_{\text{ext}}^{(j)}({\bf r})$ are
described by the one-body distribution functions $f^{(j)}({\bf r},{\bf
p},t)$.  These obey the Vlasov-Landau kinetic equations (VLE),
\begin{equation}
\partial_t f^{(j)}+\dfrac{\bf p}{m_j}\cdot{\bf \nabla}_{{\bf r}} 
f^{(j)}-
{\bf \nabla}_{{\bf r}} U^{(j)}\cdot {\bf \nabla}_{{\bf p}}f^{(j)}=
C[f^{(F)},f^{(B)}]
\label{vlasov}
\end{equation}
where the Hartree-Fock effective potential is $U^{(j)}({\bf
r},t)\equiv V_{\text{ext}}^{(j)} ({\bf r})+g n^{(\overline{j})}({\bf
r},t)$ with $\overline{j}$ denoting the species different from $j$.
Here we have set $g=2\pi \hbar^2 a/m_r$ with $a$ being the $s$-wave
scattering length of a fermion-boson pair and $m_r$ their reduced
mass, and $n^{(j)}({\bf r},t)$ is the spatial density given by
integration of $f^{(j)}({\bf r},{\bf p},t)$ over momentum degrees of
freedom.  Since we deal with low concentrations of impurities we have
neglected impurity-impurity interactions. In addition, collisions
between spin-polarized fermions are negligible at low temperature and
thus the collision integral $C$ in Eq.\ (\ref{vlasov}) involves only
collisions between fermions and impurities. This is given by
\begin{widetext}
\begin{equation}
C = \frac{\sigma}{4\pi(2\pi\hbar)^3}\int\!\! d^3p_2\,d\Omega_f
\,v\,[(1-f^{(F)})(1+f_2^{(B)})f_3^{(F)} f_4^{(B)}-
f^{(F)}f^{(B)}_2(1-f_3^{(F)}) (1+f_4^{(B)})],
\label{integralcoll}
\end{equation}
\end{widetext}
where $f^{(j)}\equiv f^{(j)}({\bf r},{\bf p},t)$ and $f_i^{(j)}\equiv
f^{(j)}({\bf r},{\bf p}_i,t)$, $d\Omega_f$ is the element of solid
angle for the outgoing relative momentum $\mathbf{p}_3-\mathbf{p}_4$,
$v=|\mathbf{v}-\mathbf{v}_2|$ is the relative velocity of the incoming
particles, and $\sigma = 4\pi a^2$ is the scattering
cross-section. The collision satisfies conservation of momentum
($\mathbf{p}+\mathbf{p}_2 = \mathbf{p}_3 + \mathbf{p}_4$) and energy
($\varepsilon + \varepsilon_2 = \varepsilon_3 + \varepsilon_4$), with
$\varepsilon_j= p_j^2/2m_j + U^{(j)}$.

Hereafter we shall focus on a specific mixture of experimental
relevance, namely, a $^{40}$K-$^{87}$Rb gas with strongly attractive
scattering length $a=-410$ Bohr radii.  At this bare scattering length
the mixture is not driven to collapse for the values of the atomic
densities that we consider in this work~\cite{Modugno2002a}.  We shall
focus on the case of isotropic traps with trap frequencies
$\omega_F\neq\omega_B$, chosen as the geometric average of the
experimental ones~\cite{Ferlaino2003a}, {\it i.e.}
$\omega_F=2\pi\times 134\,{\rm s}^{-1}$ and $\omega_B=2\pi\times
91.2\, {\rm s}^{-1}$.

\section{The collision rate}
\label{sec:collrate}

The physical observable that identifies the dynamical regime of the
gas is the quantum collision rate $\Gamma_q$. In the collisionless
regime the atoms collide only rarely and we have $\Gamma_q \ll
\omega_j$, while if $\Gamma_q \gg \omega_j$ the gas is in a
collisional regime which can be well described by hydrodynamic
equations. In the general case a kinetic treatment is necessary such as
the one used in this work.  The collision rate can then be evaluated
either from a numerical simulation which actually counts the number of
collisions at each time step, or by direct integration of the
collision integral $C$ over momenta. Numerical simulation runs for the
problem at hand will be presented in Sec.\ \ref{sec:results}.

In the second type of approach that we have mentioned above, the
starting point is the usual assumption that before a collision the
distribution functions for the incoming particles and for the
occupancy of the final states are the equilibrium Fermi and Bose
distributions. The local collision rate $\Gamma_q^{\text{loc}}$ is
then given by
\begin{widetext}
\begin{equation}
\Gamma_q^{\text{loc}} = \frac{a^2m_r^3}{(2\pi)^4\hbar^6}\int {\rm
d}^3P_i\int {\rm d}v_i v_i^3 \int {\rm d}\!\cos\theta_i \int {\rm
d}\!\cos\theta_f f_1^{(F)}f_2^{(B)}(1-f_3^{(F)})(1+f_4^{(B)})\,.
\label{eq:Gamma_loc}
\end{equation}
\end{widetext}
Here ${\bf P}_i$ and ${\bf v}_i$ are the total momentum and the
relative velocity of the incoming particles, and $\theta_i$
($\theta_f$) is the direction of the relative velocity of the incoming
(outgoing) particles in the scattering plane.  At equilibrium the
Fermi and Bose distributions in Eq.~(\ref{eq:Gamma_loc}) are given by
\begin{equation}
f_i^{(j)}=\{\exp[\beta(\varepsilon_i^{(j)}- \mu^{(j)})] +\xi\}^{-1}
\label{equilibrium}
\end{equation}
where $\xi=1$ or $-1$ respectively for fermions and bosons,
$\mu^{(j)}$ are the chemical potentials, and $\beta=1/k_BT$. The total
number of instantaneous collisions occurring in the system per unit
time is then calculated as $\Gamma_q = \int
\Gamma_q^{\text{loc}}\,d^3r$.

One can further develop the evaluation of Eq.\ (\ref{eq:Gamma_loc}) in
two limiting cases (see Appendix).  In the first case the temperature
$T$ of the fermions is very low - more precisely, $T_c \ll T \ll T_F$
where $T_F=(6N_F)^{1/3} \hbar\omega_F/k_B$ is the Fermi temperature
and $T_c=0.94 N_B^{1/3}\hbar\omega_B/k_B$ is the critical temperature
for Bose-Einstein condensation, with $N_F$ and $N_B$ being the numbers
of particles of the two species. In Eq.\ (\ref{eq:Gamma_loc}) we can
then replace $f^{(F)}$ by the zero-temperature Fermi distribution and
$f^{(B)}$ by a classical distribution of the form
\begin{equation}
f^{(B)}(\mathbf{r},\mathbf{p}) =
\frac{1}{\pi^{3/2}\,p_0^3}\rho_B(\mathbf{r})\,e^{-p^2/p_0^2}
\label{ec:classicbosons}
\end{equation}
where $\rho_B(\mathbf{r})$ is the local density and $p_0$ the momentum
spread of the impurities.

The parameter $p_0$ is crucial in determining the scattering rate in
this situation. When the momentum spread of the impurities is low,
{\em i.e.} for $p_0/p_F \ll 1$ where $p_F=\sqrt{2\,m_F\,\mu^{(F)}}$,
all collisions are forbidden by the Pauli principle since any
fermionic final state which would be allowed by kinematics belongs to
the Fermi sphere. Collisions involving empty fermionic final states
become possible on increasing $p_0$ and the collision rate starts
increasing. At the same time the volume occupied by the impurities in
the trapped system also increases and $\Gamma_q$ reaches a maximum
value when the two clouds have approximately the same size, {\em i.e.}
for $p_0\sim m_B\,\omega_B\,p_F/(m_F\omega_F)$. On an
additional increase of $p_0$ the effective number of impurities that
can interact with the fermions goes down and consequently the number
of collisions decreases. In this latter limit $p_0/p_F \gg 1$, the
leading term of the local collision rate is calculated in the Appendix
and leads to the result
\begin{equation}
\Gamma_q \simeq \dfrac{8}{\pi}a^2\dfrac{m_B^2\omega_B^3}{p_0^2}N_B
N_F\,.
\label{p0grande}
\end{equation} 
A more direct estimate in the same limit is obtained by a classical
evaluation of the collision rate through the relation
$\Gamma_q\sim\sigma v_i N_F \bar{N}_B$ with the impurity density
$\bar{N}_B \approx N_B (m_B\omega_B/p_0)^3/(4\pi/3)$. The relative
velocity $v_i$ can be set to $p_0/m_B$ and we get
\begin{equation}
\Gamma_q\simeq 
3a^2\dfrac{m_B^2\omega_B^3}{p_0^2}N_B N_F\,,
\label{p0grandeocchio}
\end{equation}
which is in good agreement with Eq.~(\ref{p0grande}). Equations
(\ref{p0grande}) and (\ref{p0grandeocchio}) show that the collision
rate rises for heavier impurities, thus favoring a
collisionless-to-collisional dynamical transition. One may consider
using in this context not only atomic impurities but also {\em e.g.}
heavier stable molecules.

The second limiting case in which progress can be made by analytical
means refers to a gas at high temperature, it is worth noticing that
in this case there would be no difference in considering as impurities
either bosons or fermions. The behavior of the mixture approaches that
of a classical two-component fluid, where the number $\Gamma_{cl}$ of
collisions per unit time is
\begin{equation}
\Gamma_{cl}= \frac{\sigma}{\pi^2}
\frac{k_r^{3/2}}{m_r^{1/2}}\,N_B\,N_F\,\beta
\label{Qclass}
\end{equation} 
(see Appendix). In Eq.\ (\ref{Qclass}) $k_r=m_B\omega_B^2\,
m_F\omega_F^2 / (m_B\omega_B^2+m_F\omega_F^2)$ is the reduced
oscillator strength.

Finally, in close-to-equilibrium situations the evolution of the
fermions can be obtained from a first-order approximation on the
collision integral as
\begin{equation}
C = -\gamma\,(f^{(F)} - f^{(F)}_{0})
\label{eq:firstorder}
\end{equation}
where $f^{(F)}_{0}$ is the equilibrium distribution. The quantity
$\gamma$ determines the damping rate of oscillatory motions of the
fermions and at low collisionality an order-of-magnitude estimate of
the damping can be obtained by evaluating one of the terms entering
the full collision integral (see {\em e.g.} \cite{Huang}). This yields
a relationship between collision rate and damping rate as 
\begin{equation}
\gamma \simeq \Gamma_q/N_F .
\label{approxgamma}
\end{equation}
The accuracy of this relation will be explored numerically in the next
Section.

\section{Dynamics of the fermions}
\label{sec:results}

The analysis of the dynamics of the mixture requires the solution of
the VLE in Eq.\ (\ref{vlasov}).  The solution is obtained by using a
fully three-dimensional concurrent code with numerical procedures
which are essentially the same as those described in
Refs. \cite{Toschi2003a, Toschi2003c} for a two-component Fermi gas,
except for the choice of the initial phase-space distributions and for
the bosonic enhancement factors entering the collision integral. The
details have already been given elsewhere \cite{Toschi2003c}.

We shall analyze two different experiments. In the first, the mixture
is prepared in an equilibrium state with distribution functions as
given in Eq.\ (\ref{equilibrium}). In the second experiment the
fermionic and bosonic clouds are prepared independently at two
different temperatures ($T$ and $T_B$) and are then superposed in
space. In this case the fermions obey a phase-space distribution given
by Eq. (\ref{equilibrium}) with a vanishing mean-field term, while the
impurities are distributed according to Eq.\ (\ref{ec:classicbosons})
with $p_0=(2m_B T_B)^{1/2}$.

\subsection{Dynamics near to equilibrium} 

Having prepared the mixture in an equilibrium state inside isotropic
traps, the system is set off by giving a slight displacement to the
center of the fermion trap. We count the collisions made per unit time
and per fermion, and report its average during roughly 8
oscillations. Therefore, at variance from what was discussed in
Section \ref{sec:collrate}, the collision rate calculated in this way
includes both the effects of the instantaneous collisions and of the
dynamics.

\begin{figure}
\includegraphics[width=\columnwidth,clip=true]{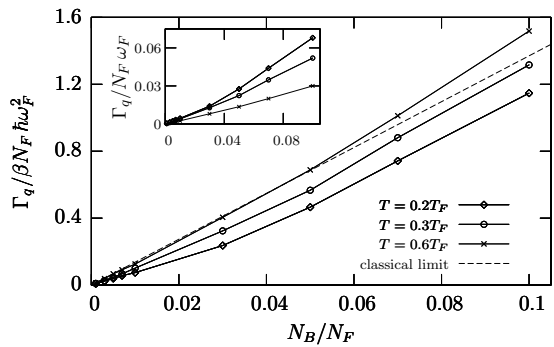}
\caption{\label{fig:GammaQ_vs_NB} Scaled collision rate
$\Gamma_q/\beta N_F$ per fermion (in units of $\hbar \omega_F^2$) in a
$^{40}$K-$^{87}$Rb mixture in isotropic confinement with $N_F=10^5$,
as a function of the impurity concentration $N_B/N_F$ at three values
of the temperature $T/T_F$ (symbols). The dashed line shows the
classical limit of Eq.~(\ref{Qclass}). The inset shows the collision
rate $\Gamma_q/N_F$ per fermion, in units of $\omega_F$.}
\end{figure}

In Fig. \ref{fig:GammaQ_vs_NB} we report the scaled collision rate
$\Gamma_q/\beta$ and (in the inset) the collision rate $\Gamma_q$ as
functions of $N_B/N_F$ at $N_F = 10^5$, for three values of the
temperature below the Fermi temperature. We are dealing in these cases
with almost classical impurities ($T/T_c \simeq 1.2 - 16$) giving a
very small contribution to the mean-field potential. The
collisionality of the gas obviously decreases with $N_B$ and goes
linearly to zero for vanishing $N_B$. The effect of cooling the gas is
instead more subtle. In spite of Pauli blocking the collision rate
\emph{increases} on cooling (see inset), as a result of the increase of
the density of impurities in the overlap region of the two
clouds. This effect has the same origin as the factor $\beta$ entering
the classical collision rate in Eq. (\ref{Qclass}) and indeed the
scaled quantity $\Gamma_q/\beta$ decreases on cooling, as is shown in
the main body of Fig.\ \ref{fig:GammaQ_vs_NB}. This also shows that
the classical formula in Eq.\ (\ref{Qclass}) gives a good account of
our results at $T=0.6\,T_F$ (see also Ref. \cite{Succi2003a}). Similar
trends are obtained in simulation runs at $N_F= 10^4$.

\begin{figure}
\includegraphics[width=\columnwidth,clip=true]{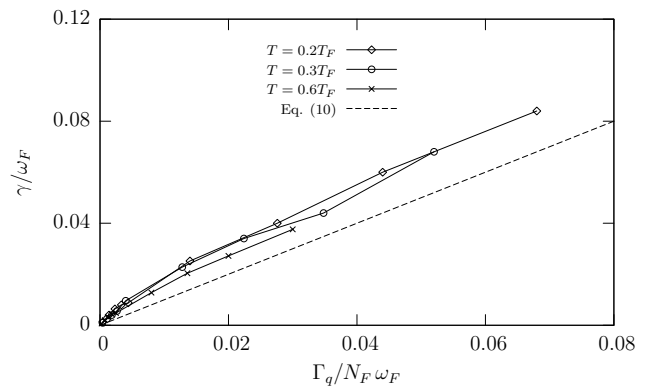}
\caption{Damping rate $\gamma$ (in units of $\omega_F$) in a
$^{40}$K-$^{87}$Rb mixture in isotropic confinement with $N_F=10^5$,
as a function of the collisionality $\Gamma_q/N_F \omega_F$ per
fermion (cf. Fig.\ \ref{fig:GammaQ_vs_NB}) and at three values of $T$
(symbols). The dashed line shows the first-order approximation given
in Eq.\ (\ref{approxgamma}).}
\label{fig:gamma_vs_Gamma}
\end{figure}

We have also analyzed the damping rate $\gamma$ of the oscillations of
the fermion cloud under the influence of the impurities and its
relationship with the collision rate.  With the present system
parameters the gas is close to the collisionless regime, as is
demonstrated by displaying in Fig. \ref{fig:gamma_vs_Gamma} the
relationship between $\gamma$ and $\Gamma_q$ for isotropic
confinement. The damping rate was obtained from the simulation data by
fitting the center-of-mass coordinate of the fermions to the
expression $z_F(t) \propto \cos (\omega_F\,t) \exp (-\gamma\,t)$. It
is seen from Fig.\ \ref{fig:gamma_vs_Gamma} that the first-order
estimate given in Eq.\ (\ref{approxgamma}) for a fluid at low
collisionality is not far from our numerical results. For what
concerns the dependence of $\gamma$ on temperature, we see from Fig.\
\ref{fig:gamma_vs_Gamma} that it is closely related to that of
$\Gamma_q$ since in this type of plot the data for different
temperatures are essentially superposed.

\subsection{Dynamics far from equilibrium}
We focus on an experiment in which, instead of varying the number of
impurities or the shape of the trap, we can control the distribution
of the impurities. In practice this could be realized by preparing
spatially separated clouds and then rapidly transferring them into the
same spatial region. This situation is closer to the system discussed
in Ref.\ \cite{Amoruso1999a}, where strongly out-of-equilibrium
collisions were assumed. Being an out-of-equilibrium situation we
choose to characterize the dynamics by displaying the time-averaged
collision rate of the system as function of $p_0$.

\begin{figure}
\includegraphics[width=\columnwidth,clip=true]{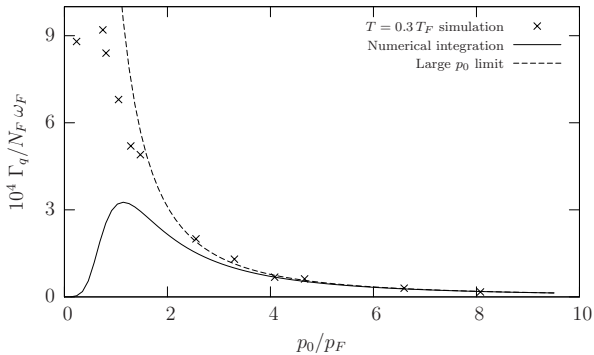}
\caption{\label{fig:Gamma_p0}Collision rate per fermion (in units of
$\omega_F$) as a function of the momentum spread $p_0$ (in units of
the Fermi momentum $p_F$) for $N_F=10^4$ $^{40}$K atoms and $N_B=100$
$^{87}$Rb impurities inside spherical traps.}
\end{figure}

The collision rate per fermion as a function of the momentum spread of
the impurities is shown in Fig.\ \ref{fig:Gamma_p0}. Here the
time-averaged results from simulation runs on a Fermi gas at $T =
0.3\,T_F$ are compared with those from the numerical integration of
Eq. (\ref{eq:Gamma_loc}) for a fully degenerate Fermi gas ($T = 0$) at
equilibrium and from its analytical expansion at large $p_0$ as given
in Eq.\ (\ref{p0grande}). Even though there are quantitative
differences between the results at $T=0$ and $T=0.3\,T_F$, the
qualitative behavior is the same at both temperatures and the
discrepancies could be attributed to the enhancement of the
collisionality at higher $T$ and to out-of-equilibrium effects missing
in Eq.\ (\ref{eq:Gamma_loc}). The peak near $p_0\simeq p_F$
corresponds to the situation where the two clouds essentially occupy
the same spatial region. Correspondingly, in view of Eq.\
(\ref{approxgamma}) also the damping rate will have a peak at the same
point.  On further increasing the momentum spread $p_0$ the effective
number of impurities that interact with the fermions diminishes and
consequently the collision and damping rates decrease.

\begin{figure}
\includegraphics[width=\columnwidth,clip=true]{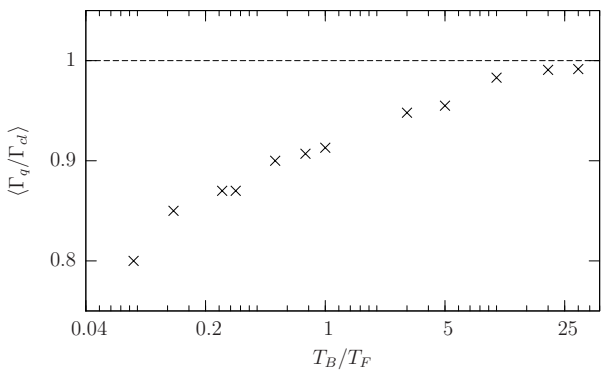} 
\caption{\label{fig:supression} Time-averaged ratio
$\langle\Gamma_q/\Gamma_{cl}\rangle$ between the quantum and the
classical collision rate at $T=0.3\,T_F$, obtained in the numerical simulation as a
function of $T_B/T_F$ (in log scale) for the gas shown in Fig.\
\ref{fig:Gamma_p0}.}
\end{figure}

Figure \ref{fig:supression} shows the time-averaged ratio $\langle
\Gamma_q/\Gamma_{cl}\rangle$ between the quantum and the classical
collision rates as a function of $T_B/T_F$, as estimated during the
simulation in correspondence to the data points in Fig.\
\ref{fig:Gamma_p0}. Almost all collisions allowed by kinematics for
$p_0>p_F$ involve empty fermionic final states so that Pauli blocking
is ineffective. For $p_0<p_F$ instead, only impurities in the tail of
the momentum distribution can contribute to the collision rate and the
Pauli suppression increases dramatically. The quantum suppression is
still lower than in the case of a two-component mixture of fermions,
where both components are subject to Pauli blocking. In this case
$\langle \Gamma_q/\Gamma_{cl}\rangle$ has been estimated to be roughly
0.65 at $T = 0.3\,T_F$ \cite{Succi2003a}, while in the present case of
bosonic impurities we find $\langle\Gamma_q/\Gamma_{cl}\rangle\simeq
0.9$ for $T_B = 0.3\,T_F$ (see Fig. \ref{fig:supression})

\section{Summary and concluding remarks}
\label{sec:remarks}

We have studied the collisionless properties of a spin-polarized Fermi
gas interacting with a small number of thermal bosonic impurities and
have focused on two related aspects of its collisionality, {\em i.e.}
the collision rate and the damping rate of oscillations. 
While the superfluidity of Bose-condensed atoms lowers the
collisionality of Bose-Fermi mixtures and the Pauli principle
operating in a second fermionic component blocks collisional events in
Fermi-Fermi mixtures, the use of thermal bosons circumvents these
limitations and may offer a feasible method to increase the
collisionality of a spin-polarized fermion gas.

For near-to-equilibrium dynamics in $^{40}$K-$^{87}$Rb mixtures
similar to those used in actual experiments at LENS
\cite{Ferlaino2003a} we have found that collisions are rare and that
the gas is close to the collisionless regime. We have suggested that
some parameters characterizing the system could be tuned to increase
its collisionality and thus drive the system towards the collisional
regime. We have also shown how in a far-from-equilibrium experiment on
a fermion gas at $T\simeq 0$ the addition of bosonic impurities with
momentum spread around the Fermi momentum $p_F$ could induce an
enhancement of collisionality. Due to the dependence of the collision
rate on the mass of the impurities, a further increase of
collisionality could be achieved by choosing as impurities heavy
particles such as $^{133}$Cs \cite{Weber2003a} and $^{172}$Yb
\cite{Takasu2003a} atoms, or strongly bound molecules.

Finally, the trap anisotropy could be raised so that for cigar-shaped
traps the axial oscillations would be damped more rapidly than the
radial ones on their own time scale. The system could thus be driven
into the intermediate collision regime on increasing the anisotropy of
the trap. The transition from the collisionless to the collisional
regime as driven by anisotropy will be investigated in detail in
future work \cite{Noi}.

\acknowledgments

This work has been partially supported by an Advanced Research
Initiative of Scuola Normale Superiore di Pisa and by the Istituto
Nazionale di Fisica della Materia within the Advanced Research Project
``Photonmatter'' and the initiative ``Calcolo Parallelo''.

\appendix

\section{Some analytical results}
\subsection{High temperature}

At high temperature the local collision rate in a two-component
mixture with only inter-species interactions is given by the classical
expression
\begin{equation}
\Gamma_{cl}^{\text{loc}}(\mathbf{r}) = \sigma \int d^3p_1\,d^3p_2\,
v\,f^{(1)}(\mathbf{r},\mathbf{p}_1) \,
f^{(2)}(\mathbf{r},\mathbf{p}_2)
\label{eq:Qclass_local}
\end{equation}
where $v$ is the relative velocity of particles 1 and 2 and the
classical distribution functions in an anisotropic trap are
\begin{equation}
f^{(j)}(\mathbf{r},\mathbf{p}) =
\frac{N_j}{(2\pi)^3}\,\beta^3\,\omega_j^2\,e^{-\beta\left(\frac{p^2}{2m_j}
+ U^{(j)}(r)\right)}.
\end{equation}
Integration of Eq.\ (\ref{eq:Qclass_local}) is straightforward and
gives
\begin{widetext}
\begin{equation}
\Gamma_{cl}^{\text{loc}}(r) = \dfrac{1}{2\,\sqrt{2}\,\pi^{7/2}}
\frac{\sigma\,\beta^{5/2}}{m_r^{1/2}} (m_1m_2)^{3/2} N_1 N_2 (\omega_1
\omega_2)^3 e^{-\beta\left(V_{\text{ext}}^{(1)}(r) +
V_{\text{ext}}^{(2)}(r)\right)}
\label{eq:Qclass_local2}
\end{equation}
\end{widetext}
if mean-field potentials can be neglected.  The total collision rate
can then be calculated by evaluating the integral of Eq.\
(\ref{eq:Qclass_local2}) over space. This yields Eq.\ 
(\ref{Qclass}) for the classical collision rate.

\subsection{Low temperature and large momentum spread of bosonic impurities}
The collision rate can also be analytically estimated when the
fermions are at $T\simeq 0$ and for large values of the momentum
spread $p_0$. In this case the fermionic distribution is the
Fermi-Dirac step function, while the impurities are taken thermally
distributed according to Eq.\ (\ref{ec:classicbosons}). Neglecting the
Bose enhancement factor and mean-field effects,
$\Gamma_q^{\text{loc}}$ can be written as
\begin{widetext}
\begin{equation}
\begin{array}{ll}
\Gamma_q^{\text{loc}}(r)&=a^2\dfrac{m_1m_2}{2m_r}\dfrac{\rho_B(r)}{\pi^{3/2}
\hbar^3p_0} \int\limits_{-\infty}^{\infty}
dP_i\,e^{-\frac{m_r^2}{m_1^2}\frac{P_i^2}{p_0^2}}\,
\int\limits_{v_{-}}^{v_{+}}dv_iv_i\,e^{-\frac{m_r^2v_i^2}{p_0^2}}
\left[e^{\frac{m_2}{m_1}\frac{p_F^2}{p_0^2}}\,e^{-\frac{m_r^2}{m_1m_2}
\left(\frac{P_i^2+m_2^2v_i^2}{p_0^2}\right)}-e^{-2\frac{m_r^2}{m_1}
\frac{P_iv_i}{p_0^2}}\right]\\&\,\times\left[-p_F^2+\dfrac{m_r^2}{m_2^2}(P_i+m_2v_i)^2
\right]\,\\
\end{array}
\label{pablo2}
\end{equation}
\end{widetext}
where the integration bounds in $v_i$, {\em i.e.}
$v_-=|P_i/m_2-p_F/m_r|$ and $v_+=(P_i/m_2+p_F/m_r)$, come from
requiring that the final states in a collision are unoccupied.

To obtain the number of collisions per unit time we integrate the
above expression inside the Thomas-Fermi radius $R_{\text{TF}}$ of the
fermion cloud. This is defined as the classical turning point of a
fermion with energy $\mu^{(F)}$ in the external trapping potential,
{\em i.e.}
\begin{equation}
R_{\text{TF}} = \sqrt{\frac{2\,\mu^{(F)}}{m_F\,\omega_F}}
\end{equation}
where we set $\mu^{(F)}$ at the non-interacting value $\mu^{(F)} =
{p_F^2}/{2m_F} = (6\,N_F)^{1/3}\,\hbar\omega_F $.  On increasing $p_0$
the integrand in Eq.\ (\ref{pablo2}) is well approximated by an
inverted parabola with zeroes at $v_{-}$ and $v_{+}$, and hence the
integral over $v_i$ is trivial. We obtain
\begin{equation}
\Gamma_{q}^{\text{loc}}(r) \simeq \frac{4}{3} \frac{a^2}{m_2}
\, \frac{\rho_B(r)}{\pi^{3/2} \hbar^3}\,p_0 p_F^3.
\label{eq:Gloc1}
\end{equation}
Expression (\ref{p0grande}) in the main text follows from taking a
local-density approximation for $p_F$ by setting $p_F
=\sqrt{2m_F(\mu^{(F)} - V_{\text{ext}}^{(F)}(r))}$ and
performing the space integration of the local collision rate in Eq.\
(\ref{eq:Gloc1}).

\end{document}